# Diagnosing the Demographic Distributions of LLM-Based Synthetic Persona Data: Reference-Distribution Dependence and Post Hoc Adjustment

Eunjeong Song · Sehee Hong
Department of Education, Korea University, Seoul, Republic of Korea

Eunjeong Song, ORCID iD: https://orcid.org/0000-0002-2302-3227
Sehee Hong, ORCID iD: https://orcid.org/0000-0001-5468-8398

## Abstract

We examine how strongly demographic-distribution diagnostics of LLM-based synthetic persona data depend on the official statistics chosen as the reference. We compared the sex × age-group × province joint distribution of the 1,000,000 Nemotron-Personas-Korea (NPK) records with Korean official statistics using total variation distance (TVD). Against the resident-registration population for April 2026, the time of use, the upper bound on the share discrepancy of any subgroup defined by the three variables was 1.81 percentage points, the margin of error of a survey with about 2,900 respondents. This value, however, depended on the reference date and population definition of the official statistics. The closest candidate examined, the 2024 census distribution for Korean nationals, gave 0.56 percentage points; no monthly resident-registration reference was closer. The reference distributions themselves also moved: between January 2025, when NPK's distance was smallest, and April 2026, the resident-registration distribution moved by more than twice that minimum distance, as did the census Korean-national distribution between 2024 and 2025. Raking reduced the range of the distance across reference months to about one fifth. Cell poststratification was applicable because none of the 204 cells was empty, and neither scheme produced extreme weights. Demographic alignment is necessary but not sufficient for response validity, and even this basic diagnostic depends on the choice of reference distribution. Providers should therefore disclose the source, reference date, population definition, and joint-cell allocation procedure of the official statistics used for demographic attributes, and users should repeat the diagnosis and adjustment against statistics current at the time of use.

**1. Introduction**

A synthetic persona is a profile of a fictitious person constructed by combining demographic and social attributes rather than a record of a real individual. A growing body of work supplies such attributes to a large language model (LLM) as conditioning information and generates simulated responses for a specified subgroup (Argyle et al., 2023; Horton, 2023). Synthetic persona data usually consist of two layers, structured attributes such as sex, age, and region, and natural-language descriptions of the person; because the two layers are produced by different procedures, they can be evaluated separately. This study addresses the first layer, the distribution of the structured demographic attributes.

Sex, age, and region are the basic auxiliary variables of sample design, stratification, weighting, and post hoc adjustment (Groves et al., 2004; Kish, 1965). Comparing their marginal and joint distributions with official statistics is therefore the most elementary quality diagnostic for synthetic persona data. Matching the sex, age, and region distributions alone does not guarantee representativeness, any more than it does for a quota sample; this diagnostic is therefore a preliminary check, not a confirmation of representativeness (Section 2.1).

Computing the distance between synthetic persona data and official statistics is one thing; interpreting the value is another. The demographic distribution of a given release of a synthetic dataset is fixed, whereas the distribution of the external population moves over time. Official statistics for the same date can also differ in how they count the population and whom they cover: the resident-registration statistics and the Population and Housing Census count people on different bases. The same dataset therefore yields different diagnostic values depending on the date and the population definition of the official statistic chosen as the reference. If only one reference distribution is used, the value obtained against it can be mistaken for a fixed property of the data.

This study addresses the problem with Nemotron-Personas-Korea (NPK; H. Kim et al., 2026), released by NVIDIA. We first fix our terminology. A *reference statistic* is an external official statistic against which NPK is compared (the resident-registration population statistics or the Population and Housing Census). A *reference distribution* is the sex × age-group × province distribution constructed from a reference statistic; it is determined by a *reference date* (which month's statistic) and a *population definition* (whether people are counted at their registered address or their usual residence, and whether foreign residents are included). The *time of use* is the most recent reference month available when a user works with the data; for this study, whose main analysis was carried out in May 2026, it is April 2026, the month of NPK's release. The distance measure and the benchmarks used to interpret its size are defined in Section 3.3.

We compare NPK's joint distribution of sex, age, and region with official statistics while varying the reference date and the population definition, and thereby assess the sensitivity of the diagnostic. Section 2.2 explains how this differs from the earlier audit of the same dataset (Bae, 2026).

We ask three research questions.

RQ1. How closely do NPK's sex, age-group, and province distributions match official statistics, and how large is the observed distance when converted into units used in survey research?

RQ2. How strongly does the observed distance depend on the choice of reference distribution, that is, on the reference date and the population definition of the reference statistic?

RQ3. How much does post hoc adjustment reduce the distance, and how does the distance that remains after adjustment depend on the reference distribution?

## 2. Background

### *2.1 Synthetic Persona Data and Survey Research*

Simulated responses generated by conditioning an LLM on synthetic personas have been discussed as a way to pretest questionnaires or to support early-stage research design (Horton, 2023; Sarstedt et al., 2024). Such data, however, involve neither a sampling process nor a real response process and cannot be equated with survey data.

Some studies report that LLMs conditioned on demographic attributes can partly reproduce the response tendencies of certain groups (Argyle et al., 2023; Horton, 2023). Others show that even when means look close to real responses, variances, relationships among variables, and sensitivity to prompts differ and stability over time is low (Bisbee et al., 2024); that the opinions of some demographic groups are insufficiently reflected (Santurkar et al., 2023); that the diversity within a group is flattened relative to real respondents (Wang et al., 2025); and that responses are sensitive to format factors such as option order and labels (Dominguez-Olmedo et al., 2024). A Korean study that simulated in-depth interview participants found that opinions were partly reproduced but personal experiences were not (C. Kim & Nah, 2024).

All of these criticisms concern the response-generation stage. This study diagnoses the earlier stage, the demographic composition. Even if the composition were exact, estimates of responses would remain biased if the response distributions within cells (groups sharing the same sex, age, and region) differed from those of the real population. Demographic alignment is therefore a necessary but not sufficient condition for response validity, and distributional diagnosis is the procedure that checks it.

### *2.2 The Dataset Under Diagnosis and Prior Work*

NPK is a publicly released synthetic persona dataset generated to reflect the demographic attributes of the Republic of Korea (H. Kim et al., 2026). NVIDIA's NeMo Data Designer pipeline builds the joint structure of the structured attributes with a probabilistic graphical model and generates natural-language descriptions with an LLM (google/gemma-4-31B-it). NPK provides sex, age, marital status, education, and occupation together with region at the province and district (si/gun/gu) levels. The dataset card states that population data from the Korean Statistical Information Service (KOSIS), among other sources, were used as seed data, but it does not identify the statistical tables, their reference dates, or their population definitions. It also states that independence assumptions were imposed on some combinations of variables because of limited public data availability, data timeliness, and the practical constraints of the probabilistic graphical model.

Bae (2026) compared NPK with official statistics and showed that marginal alignment does not guarantee joint-distribution alignment. That audit, however, examined attribute combinations that the dataset card documents as treated independently, and it fixed the reference distribution of each comparison at a single date and a single population definition. We examine the sex × age-group × province combination, for which no independence assumption is documented, and ask how much the observed distance changes as the reference date and the population definition vary.

# 3. Methods

## *3.1 Data*

### 3.1.1 Synthetic Persona Data

We analyze NPK, released by NVIDIA on Hugging Face in April 2026 under a CC BY 4.0 license. The analysis uses a fixed revision of the public repository (revision d0a9272) obtained in May 2026, and all numbers reported in this paper refer to that revision. The structured data contain 1,000,000 records, each carrying seven persona descriptions (professional, sports, arts, travel, culinary, family, and summary), which the dataset card counts as 7 million personas; throughout this paper, the unit of analysis is the record, not the persona ($N$ = 1,000,000).

We use three demographic variables: sex, age, and province. The raw data also contain a district-level (si/gun/gu) region variable, but it lacks a standard correspondence to official administrative codes, so the main analysis is restricted to the province level, where the correspondence is stable. The age range is 19 to 99, as provided by NPK. Because NPK's upper age limit is 99, the official "70 and over" category includes people aged 100 and over who do not exist in NPK; their share is about 0.02% of the population aged 19 and over, and the effect is negligible.

None of the three variables has missing values or values outside the official category systems, so all 1,000,000 records enter the analysis. Because NPK records are not real individuals, the number of records in this paper means the number of rows in the dataset, not a sample size.

### 3.1.2 Official Statistics

The resident-registration statistics were obtained through two channels. The reference distribution for the main analysis is the Ministry of the Interior and Safety's resident-registration population by administrative neighborhood (dong), sex, and single year of age as of April 30, 2026, obtained from the Public Data Portal (Ministry of the Interior and Safety, 2026a). For the reference-date analysis we used the Ministry's monthly population by age (province, sex, single year of age) for 19 reference months from April 2023 to April 2026 (Ministry of the Interior and Safety, 2026b): April 2023, October 2023, April 2024, every month from July 2024 to July 2025, October 2025, January 2026, and April 2026, so that the months outside the period of monthly coverage are spaced three to six months apart. The two sources are compiled from the same resident register, and their province × sex × age-group distributions for April 2026 coincide.

We take the resident-registration statistics as the primary reference because they are the reference that Korean survey practitioners actually use for adjustment: Korean election polls are required to compute weights from the sex, age-group, and regional composition of the electorate (National Election Survey Deliberation Commission, 2025, Articles 5 and 14), and the population used for this purpose is the Ministry of the Interior and Safety's resident-registration statistics (National Election Survey Deliberation Commission, n.d.). The electorate also includes 18-year-olds, whom NPK does not contain because its ages start at 19; the diagnosis in this paper covers ages 19 and over.

The Population and Housing Census (register-based census) compiled by the Ministry of Data and Statistics (formerly Statistics Korea) counts Korean nationals and foreign residents who have lived in Korea for three months or more as of the census date (November 1), by linking administrative records such as the resident register, the foreigner register, and university enrollment records (Ministry of Data and Statistics, 2025b). The resident-registration population is registration-based: it includes Korean nationals living abroad for study or work and excludes foreign residents. The census count of Korean nationals, in contrast, is the resident-registration population minus those not usually resident in Korea, and the census total population adds foreign residents living in Korea (Ministry of Data and Statistics, 2025b, pp. 123–124).

Table 1 summarizes the candidate reference distributions examined in this study, with their enumeration basis, coverage, and reference dates. The 2024 census (Ministry of Data and Statistics, 2025a) is the most recent register-based census that had been published at the time of use; to compare it with resident registration on the same date, we also used the resident-registration statistic of October 31, 2024, the month-end immediately preceding the census date. The 2025 census (Ministry of Data and Statistics, 2026a) was published on July 28, 2026, after the time of use (Ministry of Data and Statistics, 2026b), and was added to the analysis afterward. We therefore excluded it from the comparison that selects the candidate closest to NPK and used it only to examine whether the distance to the census also depends on the reference date (with the resident-registration statistic of October 31, 2025).

**Table 1**

*Candidate Reference Distributions Examined in This Study*

| Reference statistic | Enumeration basis | Coverage | Reference date | Population aged 19 and over |
|---|---|---|---|---|
| Resident-registration statistics (Ministry of the Interior and Safety, 2026a, 2026b) | Registered address | Korean nationals (including those living abroad; foreign residents excluded) | Last day of each month; 19 reference months from April 2023 to April 2026 (listed in Section 3.1.2) | 44,023,934 (Apr 2026), 43,901,054 (Jan 2025), 43,886,398 (Oct 2024) |
| Census, total population (Ministry of Data and Statistics, 2025a, 2026a) | Usual residence (three months or more in Korea) | Korean nationals and foreign residents living in Korea | November 1, 2024; November 1, 2025 | 44,560,450 (2024), 44,746,267 (2025) |
| Census, Korean nationals (Ministry of Data and Statistics, 2025a, 2026a) | Usual residence (three months or more in Korea) | Korean nationals (those living abroad excluded; foreign residents excluded) | November 1, 2024; November 1, 2025 | 42,633,187 (2024), 42,761,582 (2025) |

*Note.* The populations aged 19 and over are given for the three reference months used in the text. The time of use (April 2026) and the month preceding the census (October 2024) are fixed by design; January 2025 is the month identified in Section 4.2.1 as having the smallest distance, and the 2024 census Korean-national distribution is the candidate identified in Section 4.2.2 as the closest. The census counted 1,927,263 foreign residents aged 19 and over in 2024 and 1,984,685 in 2025.

### *3.2 Variables and Category Alignment*

In all sources, sex is classified as men and women. We use "age" for the variable before categorization and "age group" for a variable grouped into a particular category system. The main analysis uses six age groups: 19–29, 30–39, 40–49, 50–59, 60–69, and 70 and over. The 19–29 group is the only one that spans 11 years, because it must include age 19, the age of majority; the effect of this grouping is examined in the sensitivity analyses that refine the age groups. Provinces are unified into the 17 first-tier administrative divisions, and NPK's abbreviated province names are mapped to the official names. Because the official statistics provide counts by sex, province, and single year of age, they can be aggregated into any age grouping, converted to proportions, and compared directly with NPK's joint distribution.

Categories and cells are labeled in the order province / sex / age group (e.g., Seoul / women / 70 and over), with province names abbreviated (e.g., Seoul for Seoul Special City).

### *3.3 Analysis*

The distance between NPK and a reference distribution is computed directly from two fixed distributions and is not an estimate from a probability sample, so we do not apply significance tests. Instead, we convert the observed distance into units familiar from survey estimates and interpret it relative to two benchmarks.

Throughout, *distance* means the total variation distance between NPK and a reference distribution. The distance is a value relative to the chosen reference; it does not imply that the reference is the truth. We use *error* only for category- or cell-level differences (absolute error and maximum absolute error). A *benchmark* is a comparison standard constructed separately to gauge the size of an observed distance; benchmarks are distinct from reference distributions.

#### 3.3.1 Total Variation Distance and the Share-Discrepancy Bound

For a set of categories $C$ with reference distribution $p$ and comparison distribution $q$, the total variation distance (TVD) is half the sum of the absolute differences of the category probabilities:

$$\mathrm{TVD}(p, q) = \frac{1}{2}\sum_{c \in C} | q(c) - p(c)|.$$

We use TVD because it expresses the difference between two probability distributions on the same scale as a proportion. For any event, the difference between the probabilities that the two distributions assign to it cannot exceed the TVD (Gibbs & Su, 2002); hence for any subgroup defined by combining sex × age-group × province cells, the difference between the two sources' shares cannot exceed the TVD. TVD × 100 is therefore an upper bound, in percentage points, on the share discrepancy of any such subgroup, and the bound is attained by the subgroup consisting of all overrepresented cells.

One assumption is needed to read the share-discrepancy bound as a bound on the bias of an estimated response proportion. If, within each cell, the distribution of the variable of interest (or at least its mean) is the same in the real population and among the synthetic personas, then the difference in the means of a variable bounded in [0, 1] that arises from differences in demographic composition cannot exceed the TVD. We did not verify this assumption; if within-cell response distributions differ between the synthetic personas and the real population, the actual bias can exceed this bound.

We also report the *equivalent sample size*, the size of a simple random sample whose margin of error (for a population proportion of 0.5, at 95% confidence) equals the share-discrepancy bound. This conversion matches the bound for the worst-case subgroup to the maximum margin of error of a single proportion; for a population proportion farther from 0.5, the sample size corresponding to the same bound would be smaller. Expressing the error of a large nonprobability dataset as the size of a probability sample with sampling error of the same magnitude follows the approach of Meng (2018). The conversion serves only to compare the size of the distance with a margin of error; it does not mean that bias and sampling error are the same kind of error. The equivalent sample size is also distinct from the effective sample size used in Section 3.3.3 to summarize weight variability.

Because a small overall TVD can hide a difference concentrated in a particular category, each comparison also reports the maximum absolute error (MaxAE) together with the category or cell in which it occurs.

#### 3.3.2 Two Benchmarks

We interpret the observed distance against two benchmarks: the distance of a distribution that matches the three marginals but reproduces none of the association among the variables, and the distance that arises merely from drawing 1,000,000 records even when the generating distribution equals the official joint distribution exactly. NPK's distance is located between these two values.

The *complete-independence benchmark* is the product of the official marginal distributions of sex, age group, and province. It equals the fitted values of the loglinear model of mutual independence, in which all association among the three variables is removed (Agresti, 2013). The TVD between the official joint distribution and this benchmark indicates the strength of the association among the three variables in the population. Because the dataset card states that independence assumptions were imposed on some attribute combinations (H. Kim et al., 2026), this benchmark is not a hypothetical extreme but a design option that a generation pipeline may actually adopt. Because the unadjusted distance also contains the differences in the marginals, the comparison with this benchmark is made with the residual distance after the marginals are matched.

The *multinomial-sampling benchmark* is constructed as follows. (1) Take the official joint distribution as the true distribution. (2) Draw 1,000,000 records from it as an independent multinomial sample. (3) Compute the TVD between the sampled table and the official joint distribution. (4) Repeat steps 2–3 10,000 times, obtain the distribution of the TVD, and report its mean and 2.5th–97.5th percentiles. The benchmark for adjusted NPK is obtained by applying the same raking to each sampled table in step 3 before computing the TVD. This benchmark rests on the idealized assumption that records are drawn independently and identically; it is a yardstick for the size of a distance, not a test of the generation process.

#### 3.3.3 Post Hoc Adjustment: Cell Poststratification and Raking

Korean election-polling practice permits two weighting schemes for sex, age, and region, cell weighting and rim weighting, which correspond to cell poststratification and raking, respectively (National Election Survey Deliberation Commission, n.d.). In a Korean simulation study of election polls, the bias reduction achieved by such weighting depended on how much the weighting variables explained the variable of interest, and controlling region, sex, and age group alone barely reduced the systematic bias (Jang et al., 2014). Demographic weighting thus removes only the part of the bias that arises from differences in the distribution of the weighting variables; differences within cells remain. We apply both schemes to NPK.

*Cell poststratification* assigns to each sex × age-group × province cell a weight equal to the official share divided by the NPK share (Kalton & Flores-Cervantes, 2003). Because it uses the joint counts directly, the three-way joint distribution matches the official statistic exactly and the residual TVD is zero by construction. It cannot be applied, however, when a cell has a positive official share but no NPK records.

*Raking* adjusts one variable's marginal distribution at a time to the official marginal and iterates until all target marginals are matched (Deming & Stephan, 1940; Deville et al., 1993; Kalton & Flores-Cervantes, 2003). It requires only marginal totals, not joint counts.

Weight variability is summarized by the range of the weights, the effective sample size, and the unequal-weighting effect, that is, Kish's design effect due to unequal weighting (Kish, 1965, 1992). The effective sample size $n_{\text{eff}}$ is the square of the sum of the weights divided by the sum of the squared weights, and the unequal-weighting effect $D_w$ is the number of records divided by the effective sample size ($D_w = n/n_{\text{eff}}$, which equals $1 + \text{CV}^2$, where CV is the coefficient of variation of the weights). $D_w$ is an approximate index that summarizes only the loss of precision due to unequal weights.

#### 3.3.4 Choice of Reference Distribution

We examine how strongly the observed distance depends on the reference distribution along two axes.

The first axis is the reference date. For each of the 19 reference months of the resident-registration statistics we compute NPK's three-way TVD and the residual TVD after raking to that month's official marginals (Ministry of the Interior and Safety, 2026b). The reference month, among the 19, at which the unadjusted distance is smallest is called the *trough* (the minimum-distance month); if the minimum is shared by several months at the reported precision (four decimal places), the earliest month is used. To measure how much the population itself has moved, we also compute the TVD between the official joint distributions of the trough month and the time-of-use month. To see how the distance between two official statistics varies with the reference month, we compute the TVD between the 2024 census Korean-national distribution and each month's resident-registration distribution and present it alongside NPK's curve.

The second axis is the population definition of the reference statistic. On the 2024 reference we construct the four joint distributions—resident registration, census total population, census Korean nationals, and NPK—on the same 204 cells and compute the TVD for every pair. Among the candidates examined, the distribution closest to NPK is called the *nearest candidate*. We repeat the comparison on the 2025 reference (resident registration as of October 2025 and the 2025 census) and also compute how far the census Korean-national and total-population distributions moved between 2024 and 2025.

For each of three reference distributions—time of use, trough, and nearest candidate—we compute the age-group shares and the sum of absolute errors by age group (the sum of the absolute errors of the 34 cells in the age group; summed over the six age groups it equals twice the TVD), and for the time of use and the nearest candidate we also present cell-level errors, to see how the concentration of differences changes with the reference. Post hoc adjustment is applied to each reference distribution to check whether the distance remaining after adjustment depends on the reference. Here we add the resident-registration statistic for October 2024, the month immediately preceding the census date, so that two references that share a date but differ in population definition can be compared side by side.

The interpretation of these comparisons is limited in two ways. First, we identify the candidate reference distribution closest to NPK; we do not identify the statistic actually used in generation. Because the nearest-candidate distance is the minimum over the candidates examined, no reference examined yields a smaller distance; it is the value most favorable to NPK, and adding candidates could only lower it. Second, a distance, unlike a variance, does not decompose into source-specific components, so we do not apportion the change in distance that accompanies a change of reference among the generative model, population change over time, and differences in statistical definitions. All results are interpreted within these limits.

#### 3.3.5 Computing Environment and Sensitivity Analyses

All analyses were performed in Python 3.13 with pandas and NumPy. TVD, MaxAE, cell poststratification, and raking are deterministic; raking was judged to have converged when the maximum absolute error of the marginal proportions fell below $10^{-12}$. The multinomial-sampling benchmark uses 10,000 draws with a fixed random seed.

Three sensitivity analyses were performed: (1) splitting the 70-and-over group into 70–79 and 80 and over (238 cells); (2) refining age into 5-year groups (544 cells) and single years of age (2,754 cells), together with a check of cell coverage (the number of empty cells and the minimum cell count); and (3) replacing TVD with the Hellinger distance and the Jensen–Shannon distance. The 5-year groups begin with 19–24 and continue from 25–29 to 95–99. Analysis (2) was repeated on the age range common to NPK and the official statistics (19–99) against both the time-of-use reference and the nearest-candidate reference.

Every number, table, and figure in this paper can be regenerated from the raw data with the reproduction package, which contains the raw data, processed data, analysis code, and outputs (see Declarations).

## 4. Results

### *4.1 Distance at the Time of Use*

The three-way joint distribution of the main analysis has 204 cells (2 sexes × 6 age groups × 17 provinces). NPK's 1,000,000 records are compared with the 44,023,934 registered residents aged 19 and over.

Table 2 reports the TVD of the marginal and bivariate distributions between NPK and the resident-registration statistics at the time of use (RQ1). The TVDs for sex and province were small, whereas the TVD for age group, 0.0159, was the largest of the three marginals, with the largest error in the 70-and-over group. Among the bivariate comparisons, age group × province had the largest TVD; the differences were concentrated in the combinations that involve age.

**Table 2**

*Total Variation Distance of the Marginal and Bivariate Distributions*

| Structure | TVD | MaxAE | Category or cell with the largest error |
|---|---|---|---|
| Sex | 0.0005 | 0.0005 | — |
| Age group | 0.0159 | 0.0113 | 70 and over |
| Province | 0.0055 | 0.0031 | Gyeonggi |
| Sex × age group | 0.0159 | 0.0061 | Women / 70 and over |
| Sex × province | 0.0057 | 0.0020 | Gyeonggi / women |
| Age group × province | 0.0176 | 0.0028 | Gyeonggi / 70 and over |

*Note.* TVD = total variation distance; MaxAE = maximum absolute error. Proportions are reported to four decimal places; percentages and ratios are computed from unrounded values (this applies to all tables). Because sex is binary, the absolute errors of its two categories are identical by definition. The sex × age-group TVD equals the age-group TVD because the errors for men and women have the same sign in every age group, so no cancellation occurs when sex is collapsed.

Table 3 compares the three-way joint distribution with the two benchmarks. The TVD between the complete-independence benchmark and the official joint distribution was 0.0520, the TVD between NPK and the official joint distribution was 0.0181, and the mean of the multinomial-sampling benchmark was 0.0051. NPK's distance lies between the two benchmarks and is 3.5 times the multinomial-sampling benchmark, so the observed distance is not explained by the finite size of the dataset alone. The residual after raking, 0.0073, is the value to be compared with the complete-independence benchmark as set out in Section 3.3.2, and is one seventh of it (Section 4.3). The largest cell error was in Seoul / women / 70 and over; the largest errors clustered in the older age groups of Seoul and Gyeonggi, and no absolute error exceeded 0.15 percentage points (a proportion of 0.0015).

A TVD of 0.0181 means a share-discrepancy bound of 1.81 percentage points for subgroups defined by the three variables and an equivalent sample size of about 2,900. Reading it as a bound on the bias of a response proportion requires the assumption of equal within-cell response distributions stated in Section 3.3.1.

**Table 3**

*Three-Way Joint Distribution and the Two Benchmarks (April 2026 Reference)*

| Comparison | Three-way TVD | Cell with the largest error | Largest error |
|---|---|---|---|
| Complete-independence benchmark (product of official marginals) | 0.0520 | Gyeonggi / men / 70 and over | +0.0061 |
| NPK, unadjusted | 0.0181 | Seoul / women / 70 and over | −0.0015 |
| NPK, raked | 0.0073 | Gyeonggi / women / 19–29 | −0.0006 |
| Multinomial-sampling benchmark (mean) | 0.0051 | — | — |

*Note.* TVD = total variation distance. A positive (+) largest error indicates that the compared distribution (the benchmark or NPK) has a larger share than the official joint distribution, and a negative (−) one a smaller share. The 2.5th–97.5th percentile range of the multinomial-sampling benchmark is 0.0045–0.0057. The raked row is the time-of-use result of Section 4.3.

### *4.2 Choice of Reference Distribution*

#### 4.2.1 Reference Date

Figure 1 repeats the diagnosis against the resident-registration statistics of each of the 19 reference months (RQ2).

The unadjusted TVD ranged from 0.0075 to 0.0222 over the period examined, a threefold difference. The minimum was the same in January and February 2025 (0.0075); January 2025 is used as the trough in what follows. Over the same period the complete-independence benchmark, that is, the strength of the three-way association in the population, stayed almost constant at 0.0511–0.0520. The trough, January 2025, lies 15 months before the time of use.

The TVD at the time of use was 2.4 times the trough value. The TVD between the resident-registration joint distributions of January 2025 and April 2026 was 0.0160, which is 2.1 times the distance of 0.0075 between NPK and the January 2025 statistic. Over those 15 months, in other words, the reference distribution itself moved by more than twice NPK's minimum distance.

The dependence on the reference date is not peculiar to NPK. The triangles in Figure 1 show, independently of NPK, the distance between the 2024 census Korean-national distribution and each month's resident-registration distribution; that distance also varied more than threefold across reference months, from 0.0062 to 0.0215. Comparing a fixed distribution with resident-registration statistics of different months can move the distance by this much.

Raking to each month's official marginals left a residual TVD of 0.0067–0.0094. The range across reference months fell from 0.0147 before adjustment to 0.0027 after, about one fifth. Matching the marginals greatly reduces the variation across reference dates, but a residual joint-distribution distance remains.

**Figure 1**

*Three-Way Total Variation Distance by Reference Month: NPK and the 2024 Census Korean-National Distribution*

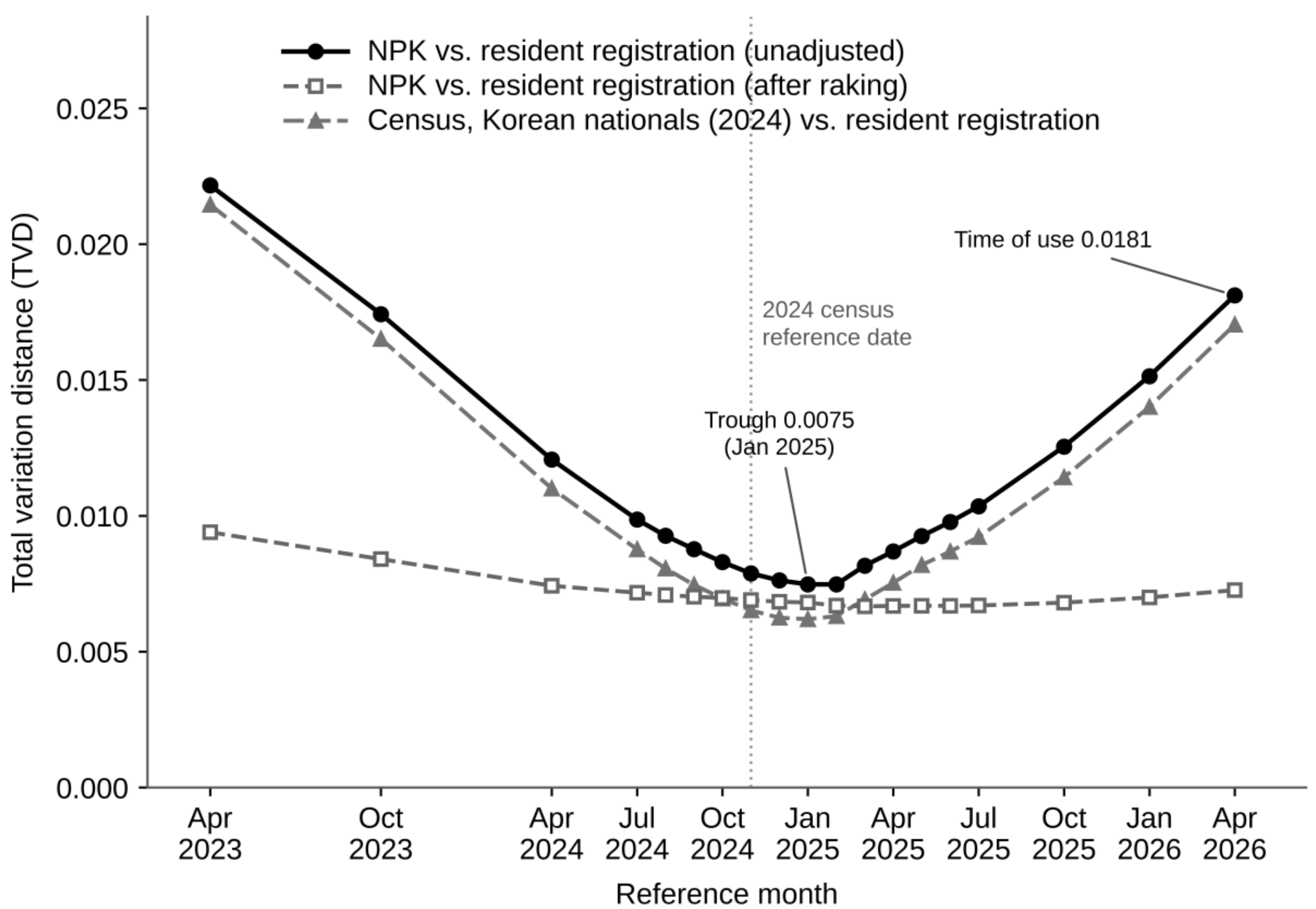

*Note.* Filled circles joined by a solid line: distance between NPK and each month's resident-registration distribution (unadjusted); open squares joined by a dashed line: residual distance after raking NPK to that month's official marginals; filled triangles joined by a dashed line: distance, independent of NPK, between the 2024 census Korean-national distribution and each month's resident-registration distribution. The vertical dotted line marks the reference date of the 2024 register-based census (November 1, 2024). The 19 reference months are April 2023, October 2023, April 2024, every month from July 2024 to July 2025, October 2025, January 2026, and April 2026; tick labels are shown for 11 of them. The annotations mark the time of use (April 2026, TVD = 0.0181) and the trough (January 2025, TVD = 0.0075; February 2025 shares the same minimum). TVD × 100 is the share-discrepancy bound, in percentage points, for subgroups defined by the three variables.

#### 4.2.2 Population Definition of the Reference Statistic

The observed distance also depended on the population definition of the reference statistic. The 2024 block of Table 4 compares, on the same 204 cells, four joint distributions: resident registration, census total population (including foreign residents), census Korean nationals, and NPK (populations aged 19 and over are given in Table 1).

**Table 4**

*Total Variation Distance Among Four Joint Distributions, 2024 and 2025 References*

| Distribution | Resident registration | Census, total population | Census, Korean nationals |
|---|---|---|---|
| **2024 reference (resident registration, Oct 2024; 2024 census)** | | | |
| Census, total population | 0.0128 | — | — |
| Census, Korean nationals | 0.0070 | 0.0131 | — |
| NPK | 0.0083 | 0.0142 | 0.0056 |
| **2025 reference (resident registration, Oct 2025; 2025 census)** | | | |
| Census, total population | 0.0135 | — | — |
| Census, Korean nationals | 0.0068 | 0.0142 | — |
| NPK | 0.0125 | 0.0149 | 0.0139 |

*Note.* Resident registration is as of October 31 and the census as of November 1 of each year. Computed on the 204 sex × age-group × province cells. Each year's matrix is symmetric, so only the lower triangle is shown. The 2025 census results were published after the time of use (Ministry of Data and Statistics, 2026b).

Four observations follow from Table 4.

First, on the 2024 reference, NPK was closest to the census Korean-national distribution (0.0056) and farthest from the census total population (0.0142). NPK was also closer to the 2024 census Korean-national distribution than to any of the 19 resident-registration months examined (0.0056 < 0.0075).

Second, the differences among the reference statistics themselves, arising from their population definitions, were not small. The distance between the census total population and the census Korean nationals was 0.0131, and the distance between resident registration and the census Korean nationals was 0.0070. Both exceed NPK's distance from the nearest candidate (0.0056).

Third, even on the same 2024 reference, the TVD between NPK and the candidate official statistics ranged from 0.0056 to 0.0142 depending on the population definition, a 2.5-fold difference. When both the reference date and the population definition were changed—from resident registration at the time of use to the 2024 census Korean nationals, the nearest candidate—the share-discrepancy bound fell from 1.81 to 0.56 percentage points and the equivalent sample size rose from about 2,900 to about 30,000. Against the nearest candidate the largest cell error was in Gyeonggi / women / 30–39 (−0.031 percentage points), and no cell had an absolute error above 0.04 percentage points.

Fourth, the distance to the census also depended on the reference date (the 2025 block of Table 4). On the 2025 reference the distance between NPK and the census Korean nationals was 0.0139, about 2.5 times the 2024 value, and among the three 2025 candidates the closest to NPK was resident registration as of October 2025 (0.0125). The census Korean-national distribution itself moved by 0.0127 between 2024 and 2025 (the total population by 0.0119), more than twice the distance of 0.0056 between NPK and the 2024 census Korean nationals. This mirrors the pattern found with the resident-registration reference.

#### 4.2.3 Age-Group and Cell-Level Errors

Table 5 gives the age-group shares and the sum of absolute errors by age group for three reference distributions, and Figure 2 shows the differences at the cell level for the time-of-use and nearest-candidate references.

Against the time-of-use reference (resident registration, April 2026), NPK's share of ages 70 and over was 1.13 percentage points below the reference, and its share of ages 19–29 was 0.81 points above. The sum of absolute errors for the 70-and-over group (1.13 percentage points) far exceeded those of the other age groups (0.33–0.83 points). When the reference statistic was held at resident registration and only the date was moved to the trough (January 2025), all six age-group shares differed by at most 0.11 percentage points, the sum of absolute errors for ages 70 and over fell to 0.21 points and was no longer the largest, and the largest group became 19–29 (0.36 points). Against the nearest candidate, the sums of absolute errors by age group were fairly even, 0.17–0.21 points. The concentration in the oldest group observed at the time of use thus changed with the reference date.

**Table 5**

*Age-Group Shares and Sums of Absolute Errors by Age Group*

| Quantity | 19–29 | 30–39 | 40–49 | 50–59 | 60–69 | 70 and over | Range |
|---|---|---|---|---|---|---|---|
| Share (%), NPK | 14.55 | 14.97 | 17.56 | 19.88 | 17.87 | 15.17 | — |
| Share (%), resident registration, Apr 2026 | 13.74 | 15.19 | 17.08 | 19.57 | 18.11 | 16.30 | — |
| Share (%), resident registration, Jan 2025 | 14.50 | 15.07 | 17.56 | 19.86 | 17.81 | 15.20 | — |
| Share (%), census Korean nationals, 2024 | 14.59 | 14.97 | 17.53 | 19.83 | 17.87 | 15.21 | — |
| Sum of absolute errors (pp), resident registration, Apr 2026 | 0.83 | 0.33 | 0.50 | 0.35 | 0.48 | 1.13 | 0.33–1.13 |
| Sum of absolute errors (pp), resident registration, Jan 2025 | 0.36 | 0.20 | 0.21 | 0.27 | 0.25 | 0.21 | 0.20–0.36 |
| Sum of absolute errors (pp), census Korean nationals, 2024 | 0.21 | 0.19 | 0.17 | 0.19 | 0.18 | 0.18 | 0.17–0.21 |

*Note.* Shares are the percentage of the population aged 19 and over in each age group. Each sum of absolute errors adds the absolute errors of the 34 cells (2 sexes × 17 provinces) in the age group; because it adds cell-level differences, it can exceed the difference in the age-group share. pp = percentage points. The three reference distributions are the time of use, the trough, and the nearest candidate, respectively; the range is the minimum and maximum of the sums of absolute errors across the six age groups. Summed over the six age groups, the sums of absolute errors equal twice TVD × 100 for that reference (3.62, 1.50, and 1.12 percentage points, respectively).

Figure 2 displays the signed errors of the 204 cells in a province × (sex × age-group) layout for the time-of-use and nearest-candidate references. Against the time-of-use reference, all four cells for ages 70 and over in Seoul and Gyeonggi were underrepresented (−0.09 to −0.15 percentage points), while the 19–29 cells for men and women in Seoul and for men in Gyeonggi were overrepresented (+0.07 to +0.09 percentage points), so the differences were concentrated among the oldest and the youngest adults of the capital region. The province-level sums of absolute errors for Gyeonggi (0.80) and Seoul (0.55) also far exceeded those of the other provinces (0.05–0.27). Against the nearest candidate, no cell had an absolute error above 0.04 percentage points and the largest province-level sum was 0.19, so this concentration disappeared.

**Figure 2**

*Cell Errors by Reference Distribution: Province × (Sex × Age Group), Time of Use and Nearest Candidate*

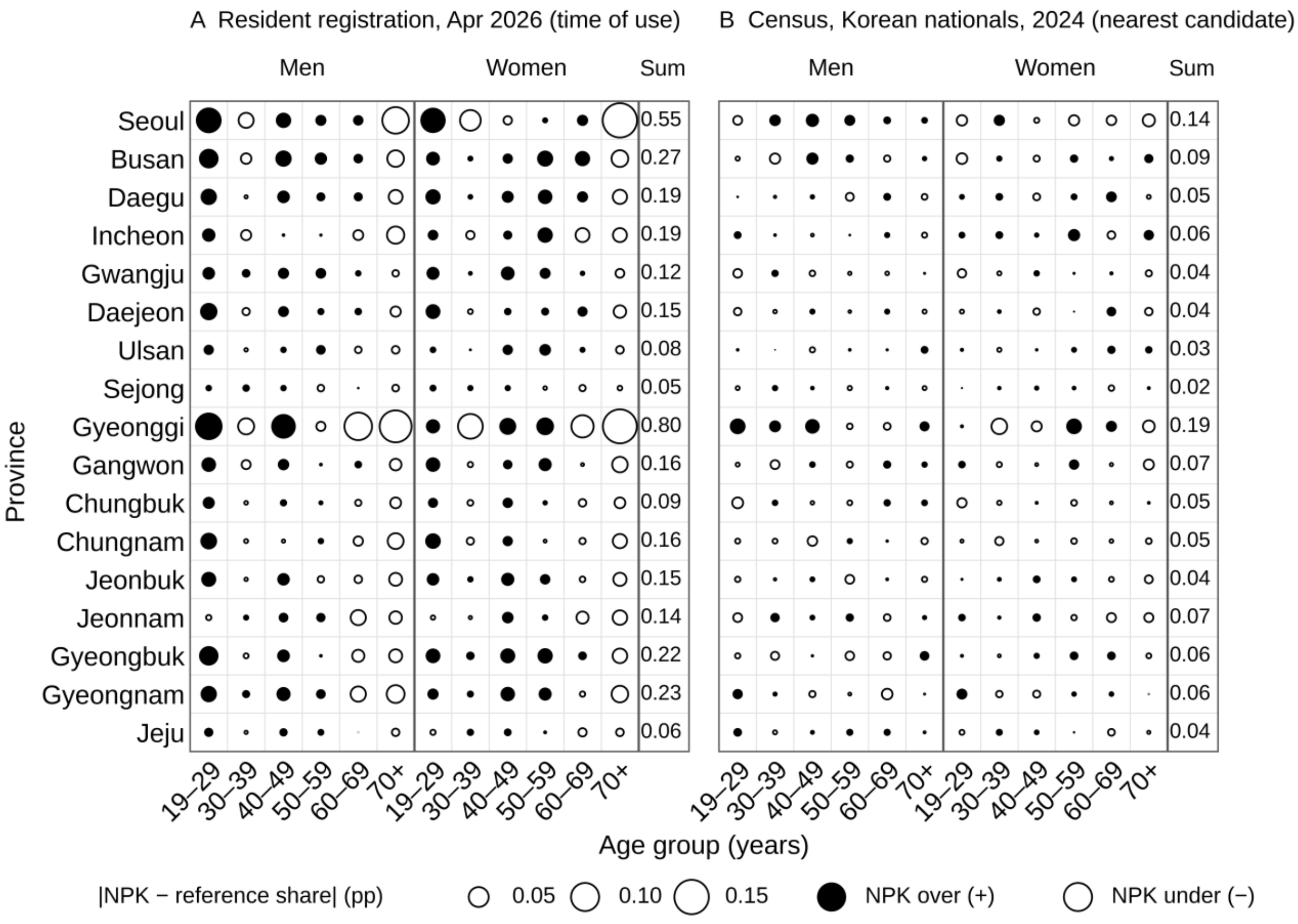

*Note.* Panel A uses the time-of-use reference (resident registration, April 2026) and Panel B the nearest-candidate reference (2024 census, Korean nationals). Rows are the 17 provinces; within each panel, the six left-hand columns are men and the six right-hand columns are women, each ordered by age group from 19–29 to 70+ (70+ denotes 70 and over). The area of each circle is proportional to the absolute difference between the NPK share and the reference share, in percentage points (legend circles: 0.05, 0.10, 0.15); filled circles indicate overrepresentation in NPK (+) and open circles underrepresentation (−). The column labeled Sum gives the sum of absolute errors of the 12 cells in each province (percentage points); summed over the 17

provinces, these equal twice TVD × 100 before rounding (3.62 percentage points in Panel A and 1.12 in Panel B). Both panels use the same scale.

### *4.3 Post Hoc Adjustment*

Table 6 reports raking and cell poststratification applied against each of four reference distributions (RQ3).

Raking to the time-of-use reference reduced the three-way TVD from 0.0181 to 0.0073. The share-discrepancy bound fell from 1.81 to 0.73 percentage points, and the equivalent sample size became about 18,000. Because none of the 204 cells was empty and the smallest cell contained 246 records, cell poststratification could be applied directly, after which the three-way TVD was zero by construction. Against the time-of-use reference the weights ranged from 0.913 to 1.093 under raking and from 0.874 to 1.235 under cell poststratification, with $D_w$ of 1.0017 and 1.0021, respectively, both close to 1. Both ranges lie within the 0.7–1.5 band that the election polling standards allow for weighting ratios by sex, age group, and region (National Election Survey Deliberation Commission, 2025, Article 5).

Because raking matches only the three marginals, a residual joint-distribution distance remains. The residual was 0.0073 against the time of use, 0.0068 against the trough, 0.0070 against resident registration for October 2024 (the month preceding the census), and 0.0054 against the nearest candidate; the distance remaining after the marginals are matched therefore also depended on the reference distribution. These residuals (0.0054–0.0073) are about one tenth to one seventh of the complete-independence benchmark (0.0520). For the three resident-registration months (October 2024, January 2025, and April 2026) the cell with the largest residual error was Gyeonggi / women / 19–29 in every case (−0.058 to −0.060 percentage points), whereas against the census Korean nationals it was Gyeonggi / women / 30–39 (−0.031 percentage points).

Against the nearest candidate, NPK's distance was 0.0056 before and 0.0054 after raking. When 1,000,000 records are drawn independently from the nearest-candidate distribution and the same raking is applied, the 95% range of the distance is 0.0045–0.0057 before and 0.0042–0.0054 after raking; both NPK values lie near the upper end of the respective range. For the joint distribution of the three variables, then, the residual distance against the nearest candidate is no larger than the upper end of the variation expected from drawing 1,000,000 records from the reference distribution itself. The time-of-use residual of 0.0073 lies well outside the same range.

**Table 6**

*Post Hoc Adjustment by Reference Distribution*

| Reference distribution | TVD before adjustment | TVD after raking | Weight range | $D_w$ |
|---|---|---|---|---|
| Resident registration, Apr 2026 (time of use) | 0.0181 | 0.0073 | 0.913–1.093 | 1.0017 |
| Resident registration, Jan 2025 (trough) | 0.0075 | 0.0068 | 0.971–1.031 | 1.0001 |
| Resident registration, Oct 2024 (month preceding the census) | 0.0083 | 0.0070 | 0.961–1.035 | 1.0002 |
| Census, Korean nationals, 2024 (nearest candidate) | 0.0056 | 0.0054 | 0.986–1.013 | 1.0000 |

*Note.* TVD = total variation distance; $D_w$ = unequal-weighting effect. "TVD after raking" is the residual three-way TVD after the three marginals are matched to the official marginals of each reference distribution; in all four conditions raking converged with a maximum absolute error of the marginal proportions below $10^{-12}$. The weight range and $D_w$ refer to raking. The TVD after cell poststratification is zero by construction in all four conditions because the joint counts are used directly. The raked multinomial-sampling benchmark, obtained by drawing 1,000,000 records independently from the reference distribution and applying the same raking, had a mean of 0.0048 and 2.5th–97.5th percentiles of 0.0042–0.0054 in all four conditions (for the unadjusted benchmark see Table 3).

The sensitivity analyses gave the following results.

First, when the oldest group was split into 70–79 and 80 and over (238 cells), the TVD was 0.0182 before and 0.0080 after raking, close to the main-analysis values (0.0181 and 0.0073), and cell poststratification remained applicable because no cell was empty ($D_w$ = 1.0023). The largest cell error was in Gyeonggi / women / 70–79 (−0.135 percentage points).

Second, refining age into 5-year groups (544 cells) and single years (2,754 cells) increased the distance. Before raking, the TVD was 0.0265 (5-year) and 0.0421 (single-year) against the time-of-use reference and 0.0086 and 0.0371 against the nearest candidate; after raking, the residual was 0.0115 and 0.0219 against the time of use and 0.0080 and 0.0211 against the nearest candidate. The 95% range of the raked multinomial-sampling benchmark was 0.0070–0.0082 at 5-year and 0.0167–0.0179 at single-year resolution, the same for both references. Against the nearest candidate, therefore, the 5-year residual lay within the benchmark range whereas the single-year residual was 1.18 times the benchmark's upper limit; against the time-of-use reference the residual exceeded the upper limit at all three resolutions, including the 10-year resolution of Table 6. Empty cells also appeared as the resolution was refined (one at 5-year and 25 at single-year resolution), together with cells containing a single record, so cell poststratification is limited under the refined conditions. The full set of figures by resolution is included in the reproduction package (see Declarations).

Third, replacing TVD with the Hellinger distance or the Jensen–Shannon distance preserved the ordering of the four quantities of the main analysis (the complete-independence benchmark, unadjusted NPK, raked NPK, and the multinomial-sampling benchmark), so the conclusions of the main analysis do not depend on the choice of distance measure.

## 5. Discussion

We examined how strongly the diagnosis of NPK's sex × age-group × province joint distribution depends on the choice of reference distribution. Against the time-of-use reference, resident registration for April 2026, the TVD was 0.0181 (a share-discrepancy bound of 1.81 percentage points); against the nearest of the candidates examined, the 2024 census Korean nationals, it was 0.0056 (0.56 percentage points). The reference distributions themselves also moved substantially: the distance the resident-registration population structure moved over 15 months (0.0160) and the distance the census Korean-national distribution moved over one year (0.0127) were each more than twice NPK's minimum distance from the corresponding statistic (0.0075 and 0.0056). The distance between a synthetic persona dataset and official statistics is therefore not a single quality indicator fixed to the data, and any reported diagnostic should state the reference date and the population definition of the reference statistic.

### *5.1 Implications for Survey Research*

First, the external demographic alignment of synthetic persona data should be rechecked against official statistics current at the time of use. Not only the distance but also the cells in which errors concentrated changed with the reference month. Compared with resident registration at the time of use, the errors concentrated in the 70-and-over cells of Seoul and Gyeonggi (Figure 2), but moving the reference month to January 2025 removed the concentration in the oldest age group (Table 5). The pattern in older-age cells seen at the time of use therefore cannot be read directly as a limitation of the generative model in reproducing older adults.

Second, the population definition of the reference statistic should be stated. Official statistics for the same period differed in their joint distributions according to whether people were counted at their registered address or their usual residence and whether foreign residents were included, and the TVD to NPK ranged from 0.0056 to 0.0142 on the 2024 reference. Even the two most recent statistics available at the time of use, resident registration for April 2026 and the 2024 census Korean nationals, give share-discrepancy bounds of 1.81 and 0.56 percentage points. Data providers should disclose, at a minimum, the source, reference date, and population definition of the statistics used for demographic attributes, together with the joint-cell structure and the allocation method. The procedure proposed here is the diagnosis that users can perform with external official statistics when this information is not disclosed or when the population has changed since the data were released. Which reference distribution to use for the diagnosis and the adjustment is determined by the population the user intends to represent; for a use such as election polling, for example, the reference is the electorate on the resident-registration basis.

Third, the role of post hoc adjustment is confined to the distributions of the variables used in the adjustment. When official joint counts are available and no cell is empty, cell poststratification matches the three-way joint distribution to the reference statistic by construction, whereas raking matches the marginals but leaves a residual distance of about 0.005–0.007 (Table 6), one tenth to one seventh of the complete-independence benchmark. Weight variability was small under both schemes ($D_w \leq 1.0021$ in the main analysis). The $D_w$ values reported here apply the weights to all 1,000,000 records; users who draw a subsample should reevaluate weight variability in that subsample. Cell poststratification does not align marital status, education, occupation, or the natural-language descriptions with the population, and at single-year resolution empty cells appeared and the residual distance exceeded the benchmark range. The resolution of the diagnosis and the adjustment should match the resolution of the subgroups actually used.

Fourth, matching the demographic joint distribution and establishing response validity are separate steps. Reading our share-discrepancy bound as a bound on response bias requires that within-cell conditional response distributions be the same for the real population and the synthetic personas, and studies comparing LLM-generated responses with those of real respondents show that this assumption may not hold. In real respondents, group differences accumulate as demographic attributes intersect, whereas in LLM-generated responses one or two attributes dominate (Rennard & Xypolopoulos, 2026), and LLMs portray the diversity within a group as flatter than it is (Wang et al., 2025). Even if the joint distribution of sex, age group, and province is matched exactly to official statistics, there is no guarantee that an LLM uses the combined attributes in the way real respondents do.

Taken together with the prior work reviewed in Section 2.1, our results suggest that synthetic persona data are best viewed not as a substitute for survey data but as a resource that supports the early stages of a survey, such as reviewing draft questions, constructing subgroup scenarios, and assembling auxiliary variables. A share-discrepancy bound of 1.81 percentage points is smaller than the margin of error of a 1,000-respondent survey (about ±3.1 percentage points) but equal to that of a survey of about 2,900. Because it is a bound, the actual bias may be smaller; still, under the assumption of Section 3.3.1, using the unadjusted data in support of a survey larger than about 2,900 respondents could leave a bias larger than that survey's sampling error. After raking the same relation holds at about 18,000 respondents (bound 0.73 percentage points), and because bias does not shrink with the number of records (Meng, 2018), the 1,000,000 records do not offset it.

### *5.2 Limitations and Future Research*

First, this is a case study of three auxiliary variables in one dataset at the province level. The magnitudes reported concern NPK; what generalizes is the diagnostic framework. Applying the same framework to Nemotron-Personas datasets for other languages (e.g., Meyer & Corneil, 2025) and to data from other generation pipelines is a task for future work. NPK's other attributes (marital status, education, occupation) and its district-level (si/gun/gu) distribution were not diagnosed because official joint counts of comparable quality were not available. The small distance observed for the joint distribution of the three variables therefore cannot be extended to other attribute combinations; for combinations with documented independence assumptions, the opposite result has been reported (Bae, 2026).

Second, because the candidate reference distributions are limited to resident registration and the 2024 and 2025 censuses (total population and Korean nationals), the nearest candidate cannot be taken to be the statistic actually used in generation. Nor could we separate, within the residual of 0.0054 against the nearest candidate, systematic differences due to the generative model from chance variation in a single realization; distinguishing the two would require generating several datasets with the same pipeline and measuring the variation among them.

Third, the response validity of the synthetic personas lies outside the scope of this study. Future research should directly test within-cell conditional response distributions, correlations among items, the factor structure and measurement invariance of scales, and the stability of repeated generation, using items identical or functionally equivalent to those of real surveys in designs that control prompt conditions, repeated generation, model versions, and demographic adjustment.

## Declarations

Data, materials, and code are available in Harvard Dataverse (https://doi.org/10.7910/DVN/BUFIGC). No funding was received for this study. The authors declare no competing interests. During preparation of this work, the authors used DeepL to translate the initial draft and improve English readability, and OpenAI Codex to assist with refinement and error checking of the research code. The authors reviewed and edited all outputs and take full responsibility for the final manuscript and research code.

## References

Agresti, A. (2013). *Categorical data analysis* (3rd ed.). Wiley.

Argyle, L. P., Busby, E. C., Fulda, N., Gubler, J. R., Rytting, C., & Wingate, D. (2023). Out of one, many: Using language models to simulate human samples. *Political Analysis, 31*(3), 337–351. https://doi.org/10.1017/pan.2023.2

Bae, J. (2026). *Marginal alignment does not guarantee joint-distribution fidelity: An official-reference audit of Nemotron-Personas-Korea with cross-locale replication*. arXiv. https://doi.org/10.48550/arXiv.2606.12433

Bisbee, J., Clinton, J. D., Dorff, C., Kenkel, B., & Larson, J. M. (2024). Synthetic replacements for human survey data? The perils of large language models. *Political Analysis, 32*(4), 401–416. https://doi.org/10.1017/pan.2024.5

Deming, W. E., & Stephan, F. F. (1940). On a least squares adjustment of a sampled frequency table when the expected marginal totals are known. *The Annals of Mathematical Statistics, 11*(4), 427–444. https://doi.org/10.1214/aoms/1177731829

Deville, J.-C., Särndal, C.-E., & Sautory, O. (1993). Generalized raking procedures in survey sampling. *Journal of the American Statistical Association, 88*(423), 1013–1020. https://doi.org/10.1080/01621459.1993.10476369

Dominguez-Olmedo, R., Hardt, M., & Mendler-Dünner, C. (2024). Questioning the survey responses of large language models. *Advances in Neural Information Processing Systems, 37*, 45850–45878. https://doi.org/10.52202/079017-1458

Gibbs, A. L., & Su, F. E. (2002). On choosing and bounding probability metrics. *International Statistical Review, 70*(3), 419–435. https://doi.org/10.1111/j.1751-5823.2002.tb00178.x
Groves, R. M., Fowler, F. J., Jr., Couper, M. P., Lepkowski, J. M., Singer, E., & Tourangeau, R. (2004). *Survey methodology*. Wiley.
Horton, J. J. (2023). *Large language models as simulated economic agents: What can we learn from* Homo silicus*?* (NBER Working Paper No. 31122). National Bureau of Economic Research. https://doi.org/10.3386/w31122
Jang, D. H., Hong, Y., & Cho, S. K. (2014). Gajung bangbeobeuro seongeo yeoronjosaui pyeonhyangeul eolmana juril su inna: Keompyuteo simyulleisyeon sarye [How much can weighting methods reduce bias in presidential election surveys: Results from a computer simulation]. *Josa Yeongu, 15*(2), 105–121.
Kalton, G., & Flores-Cervantes, I. (2003). Weighting methods. *Journal of Official Statistics, 19*(2), 81–97.
Kim, C., & Nah, K. (2024). Daehyeong eoneo model jungsimui saengseonghyeong ingongjineungeul hwaryonghan dijain riseochi: Simcheung inteobyuui simyulleisyeoneul jungsimeuro [LLM generative AI in design research: Focused on in-depth interview and its simulation]. *Hanguk Dijain Munhwa Hakhoeji, 30*(2), 65–75. https://doi.org/10.18208/ksdc.2024.30.2.65
Kim, H., Ryu, J., Lee, J., Ryu, H., Praveen, K., Prayaga, S., Thadaka, K., Jennings, W., Sadeghi, B., Sharabiani, A., Choi, Y., & Meyer, Y. (2026). *Nemotron-Personas-Korea: Synthetic personas aligned to real-world distributions for Korea* (Revision d0a9272) [Data set]. Hugging Face. https://huggingface.co/datasets/nvidia/Nemotron-Personas-Korea
Kish, L. (1965). *Survey sampling*. Wiley.
Kish, L. (1992). Weighting for unequal $P_i$. *Journal of Official Statistics, 8*(2), 183–200.
Meng, X.-L. (2018). Statistical paradises and paradoxes in big data (I): Law of large populations, big data paradox, and the 2016 US presidential election. *The Annals of Applied Statistics, 12*(2), 685–726. https://doi.org/10.1214/18-AOAS1161SF
Meyer, Y., & Corneil, D. (2025). *Nemotron-Personas-USA: Synthetic personas aligned to real-world distributions* [Data set]. Hugging Face. https://huggingface.co/datasets/nvidia/Nemotron-Personas-USA
Ministry of Data and Statistics. (2025a). *Ingu chongjosa: Yeollyeong mit seongbyeol ingu – eup-myeon-dong* [Population and Housing Census: Population by age and sex – eup/myeon/dong] (2024 census; Table DT_1IN1503, queried by province, sex, and single year of age) [Data set]. Korean Statistical Information Service. Retrieved September 14, 2026, from https://kosis.kr/statHtml/statHtml.do?orgId=101&tblId=DT_1IN1503
Ministry of Data and Statistics. (2025b, July 29). *2024-nyeon ingu jutaek chongjosa gyeolgwa (deungnok senseoseu bangsik)* [Results of the 2024 Population and Housing Census (register-based census)] [Press release]. https://mods.go.kr/board.es?act=view&bid=203&list_no=437767&mid=a10301020100
Ministry of Data and Statistics. (2026a). *Ingu chongjosa: Yeollyeong mit seongbyeol ingu – eup-myeon-dong* [Population and Housing Census: Population by age and sex – eup/myeon/dong] (2025 census; Table DT_1IN1503, queried by province, sex, and single year of age) [Data set]. Korean Statistical Information Service. Retrieved September 14, 2026, from https://kosis.kr/statHtml/statHtml.do?orgId=101&tblId=DT_1IN1503
Ministry of Data and Statistics. (2026b, July 28). *2025-nyeon ingu jutaek chongjosa gyeolgwa (deungnok senseoseu bangsik)* [Results of the 2025 Population and Housing Census (register-based census)] [Press release]. https://mods.go.kr/board.es?act=view&bid=203&list_no=446219&mainXml=Y&mid=a10301010000

Ministry of the Interior and Safety. (2026a). *Jiyeokbyeol (haengjeongdong) seongbyeol yeollyeongbyeol jumin deungnok ingusu_20260430* [Resident-registration population by administrative neighborhood (dong), sex, and age, as of April 30, 2026] (File No. 15097972) [Data set]. Public Data Portal. https://www.data.go.kr/data/15097972/fileData.do

Ministry of the Interior and Safety. (2026b). *Jumin deungnok ingu tonggye: Yeollyeongbyeol ingu hyeonhwang* [Resident-registration population statistics: Population by age] (monthly data by province, sex, and single year of age, April 2023 to April 2026) [Data set]. Retrieved July 12, 2026, from https://jumin.mois.go.kr/

National Election Survey Deliberation Commission. (n.d.). *Seongeo yeoronjosa gwallyeon yongeo haeseol: 5. Gajunggap sanchul mit jeogyong* [Glossary of election polling terms: 5. Calculation and application of weights] [Glossary panel on poll-registration detail pages]. Retrieved September 17, 2026, from https://www.nesdc.go.kr/portal/bbs/B0000005/view.do?nttId=18890&menuNo=200467

National Election Survey Deliberation Commission. (2025). *Seongeo yeoronjosa gijun* [Standards for election polls] (Notification No. 2025-1, amended December 18, 2025, effective January 1, 2026). https://www.nesdc.go.kr/portal/content/view.do?menuNo=300010

Rennard, V., & Xypolopoulos, C. (2026). *Large language models simulate intersectional synthetic identities with a budget of one to two dimensions*. arXiv. https://doi.org/10.48550/arXiv.2608.23005

Santurkar, S., Durmus, E., Ladhak, F., Lee, C., Liang, P., & Hashimoto, T. (2023). Whose opinions do language models reflect? *Proceedings of Machine Learning Research, 202*, 29971–30004. https://proceedings.mlr.press/v202/santurkar23a.html

Sarstedt, M., Adler, S. J., Rau, L., & Schmitt, B. (2024). Using large language models to generate silicon samples in consumer and marketing research: Challenges, opportunities, and guidelines. *Psychology & Marketing, 41*(6), 1254–1270. https://doi.org/10.1002/mar.21982

Wang, A., Morgenstern, J., & Dickerson, J. P. (2025). Large language models that replace human participants can harmfully misportray and flatten identity groups. *Nature Machine Intelligence, 7*(3), 400–411. https://doi.org/10.1038/s42256-025-00986-z